\documentclass[aps,prl,twocolumn,floatfix,superscriptaddress]{revtex4-1}
\pdfoutput=1
\usepackage{amsfonts,amsmath,color,graphicx,subfigure}
\usepackage[normalem]{ulem}
\usepackage{url}

\begin{document}
\title{Opinion Formation under Antagonistic Influences}
\author{Deepak Bhat}
\email{deepak.bhat@santafe.edu}
\affiliation{Santa Fe Institute, 1399 Hyde Park Rd., Santa Fe, New Mexico 87501, USA}
\author{S. Redner}
\email{redner@santafe.edu}
\affiliation{Santa Fe Institute, 1399 Hyde Park Rd., Santa Fe, New Mexico 87501, USA}

\begin{abstract}
  We study the opinion dynamics in a generalized voter model in which voters
  are additionally influenced by two antagonistic news sources, whose effect
  is to promote political polarization.  We show that, as the influence of
  the news sources is increased, the mean time to reach consensus is
  anomalously long, the time to reach a politically polarized state is quite
  short, and the steady-state opinion distribution exhibits a transition from
  a near consensus state to a politically polarized state.

\end{abstract}

\maketitle 

A disheartening feature of current social discourse is its high degree of
political polarization, particularly in the US and Europe (see, e.g.,
\cite{AG05,BG08,FA08,S10,P13,IW14,pew}).  In recent decades, this
polarization has increased to the point where, in the US, parents affiliated
with a given political party are loathe to have their children wed someone
affiliated with the other major party (see, e.g., \cite{IKT08}).  We have no
illusions of being able to explain the complex societal forces that have led
to this situation.  What we believe we can understand, however, are the
consequences of this increased polarization on the dynamics of opinion
formation.

\begin{figure}[ht]
\centerline{
\includegraphics[width=0.4\textwidth]{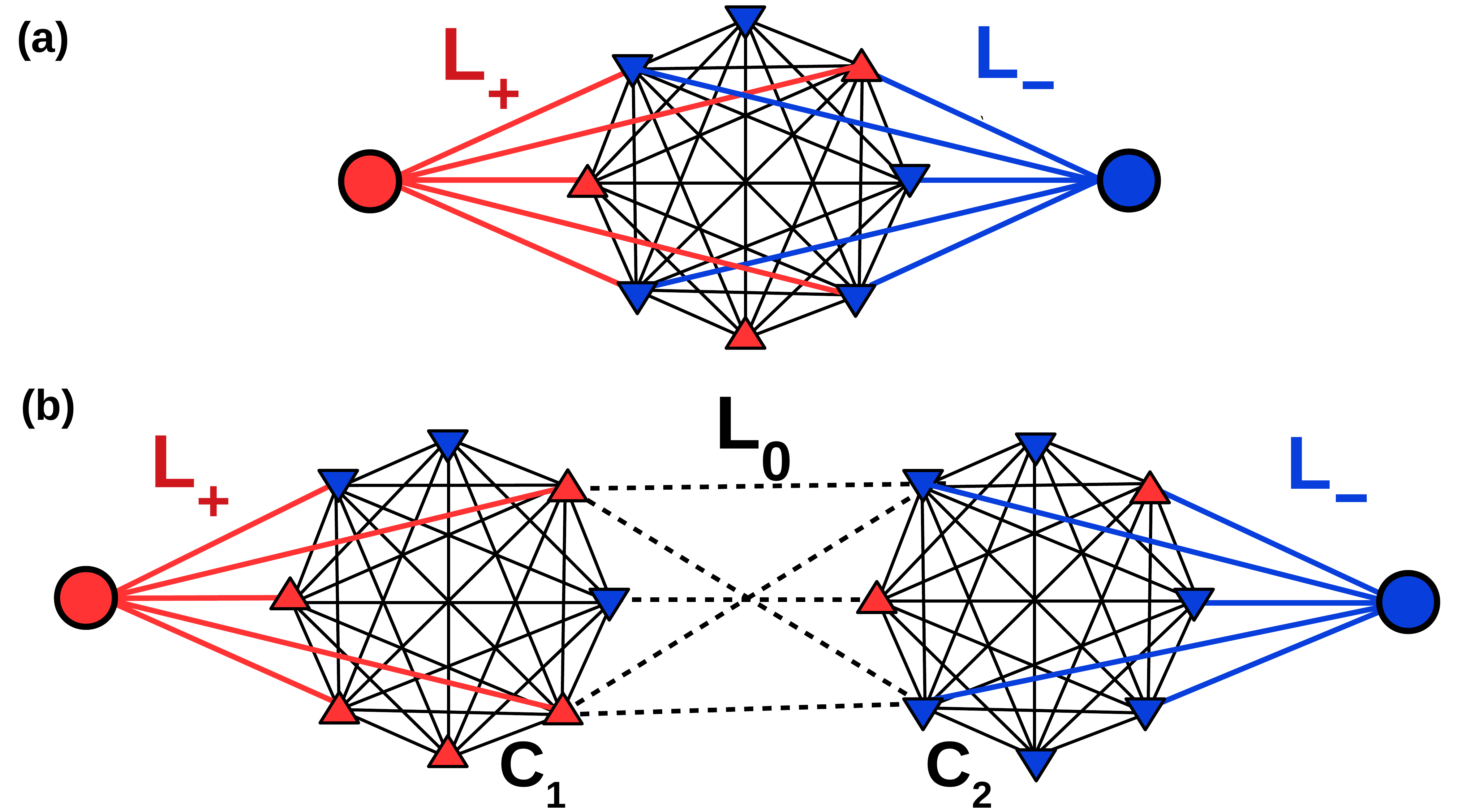}}
\caption{Two antagonistic news sources (large circles) that influence voters
  of different opinions (small up and down triangles) that are situated on
  either: (a) a complete graph, or (b) a two-clique graph.  The news sources
  have $L_\pm$ links to individuals.  For the two-clique graph, there are
  $L_0$ links between voters in different cliques.  }
\label{graphs}
\end{figure}

Our modeling is based on the framework of the voter
model~\cite{clifford1973model,holley1975ergodic,Cox,Liggett,R19} that is
augmented by the influence of competing news sources.  Many news sources
promulgate a fixed political viewpoint~\cite{IH09,L13,MY17} and news
consumers predominantly consult sources that align with their own political
persuasion.  We therefore describe a society as being influenced by two news
media sources of opposite political leanings (Fig.~\ref{graphs}).  These
sources are effectively ``zealots'' in the framework of the voter
model~\cite{M03,MG05,MPR07}, in that they perpetually maintain their
political opinion.  While many variants of the voter model---inspired by real
decision making---have been investigated (see, e.g.,
\cite{CVV03,SEM05,SR05,Sood,Castellano-q,Masuda,XSK11,Masuda2,Volovik,GSRME14}),
the role of news media has apparently not been considered (but
see~\cite{VLP19} for a study related to ours).  Our goal is to determine the
role of two news sources with opposing perspectives on the dynamics of public
opinion.

Each individual voter has two possible opinion states, denoted as $+$ and
$-$.  Individual opinions are updated according to voter model dynamics: a
randomly selected voter adopts the opinion of a randomly selected neighbor.
We account for the different propensities of news media and neighboring
voters to influence a given voter as follows: for a voter linked to one news
source and $k$ other voters, the news source is picked with probability $p/R$
and a neighboring voter is picked with probability $(1-p)k/R$, where
$R=p+k(1-p)$ is the total rate of picking any neighbor.  The parameter $p$
thus quantifies the relative influence of a news source and a neighboring
voter.  (If a voter is connected to both news sources, then $R=2p+k(1-p)$.)~
Once an interaction partner is selected, the voter adopts the opinion of this
partner.  This update step is repeated ad infinitum.

We treat two types of social networks (Fig.~\ref{graphs}): (a) A
complete graph of $N$ voters, with $L_+$ ($L_-$) connections between voters
and the $+$ $(-)$ news source.  The news sources connect either to random
voters or to disjoint voters. (b) More realistically, a two-clique graph with
$N$ voters in each clique, with $L_+$ connections between the $+$ news source
and random voters on clique $C_1$ (and correspondingly for $C_2$), and
$L_0=N^{\beta}$ links between nodes in different cliques.  In both cases,
$0< L_{\pm}\leq N$, with corresponding link densities $\ell_{\pm}=L_{\pm}/N$.

We focus on four characteristics of the collective opinion state: (i) the
consensus time $T_{\rm con}$, defined as the average time to reach either $+$ or $-$
unanimity; (ii) the polarization time $T_{\rm pol}$, defined as the average time to
go from a state with non-zero magnetization (the difference in the fraction
of $+$ and $-$ voters) to a politically polarized state of zero
magnetization; (iii) the exit probability, defined as the probability to
eventually reach $+$ consensus when the initial density of $+$ voters equals
$x$, and (iv) the steady-state opinion distribution.

Our main results are: (i) $T_{\rm con}$ typically grows algebraically with $N$, with
a non-universal exponent that can be arbitrarily large.  Based on an
annealed-link approximation to be discussed below, we find, for voters on the
complete graph:
\begin{subequations}
\label{times}
  \begin{align}
 \label{Tc-N}
T_{\rm con}  \sim 
\begin{cases}
 N &\qquad  0\leq \alpha <1\,,\\
 N\ln N &\qquad \alpha=1\,,\\
 N^{\alpha}& \qquad \alpha>1\,,
\end{cases}
\end{align}
where $\alpha=\min(\alpha_+,\alpha_-)$ and $\alpha_{\pm}=p\ell_{\pm}/(1-p)$.
For voters on the two-clique graph, in which the news sources have equal link
densities ($\ell_+=\ell_-\equiv\ell$) and the cliques are sparsely
interconnected ($\beta<1$)
\begin{align}
\label{two-clique-tc}
 T_{\rm con} \sim
 \begin{cases}
N^{2-\beta} &\qquad  0\leq \alpha< 1\,,\\
N^{2-\beta}\ln N &\qquad \alpha= 1\,,\\
N^{\alpha+1-\beta} &\qquad   \alpha>1\,.
\end{cases}
\end{align}
\end{subequations}
For $p\to 1$, i.e., influential news sources, the exponent of the consensus
time becomes arbitrarily large.  That is, competing and well-connected news
sources promote political polarization. Our results for $T_{\rm con}$ for $p\to 0$
for the two-clique graph are consistent with a previous study of the voter
model on this graph~\cite{Masuda3}.

\begin{figure}[ht]
\centerline{  \includegraphics[width=0.35\textwidth]{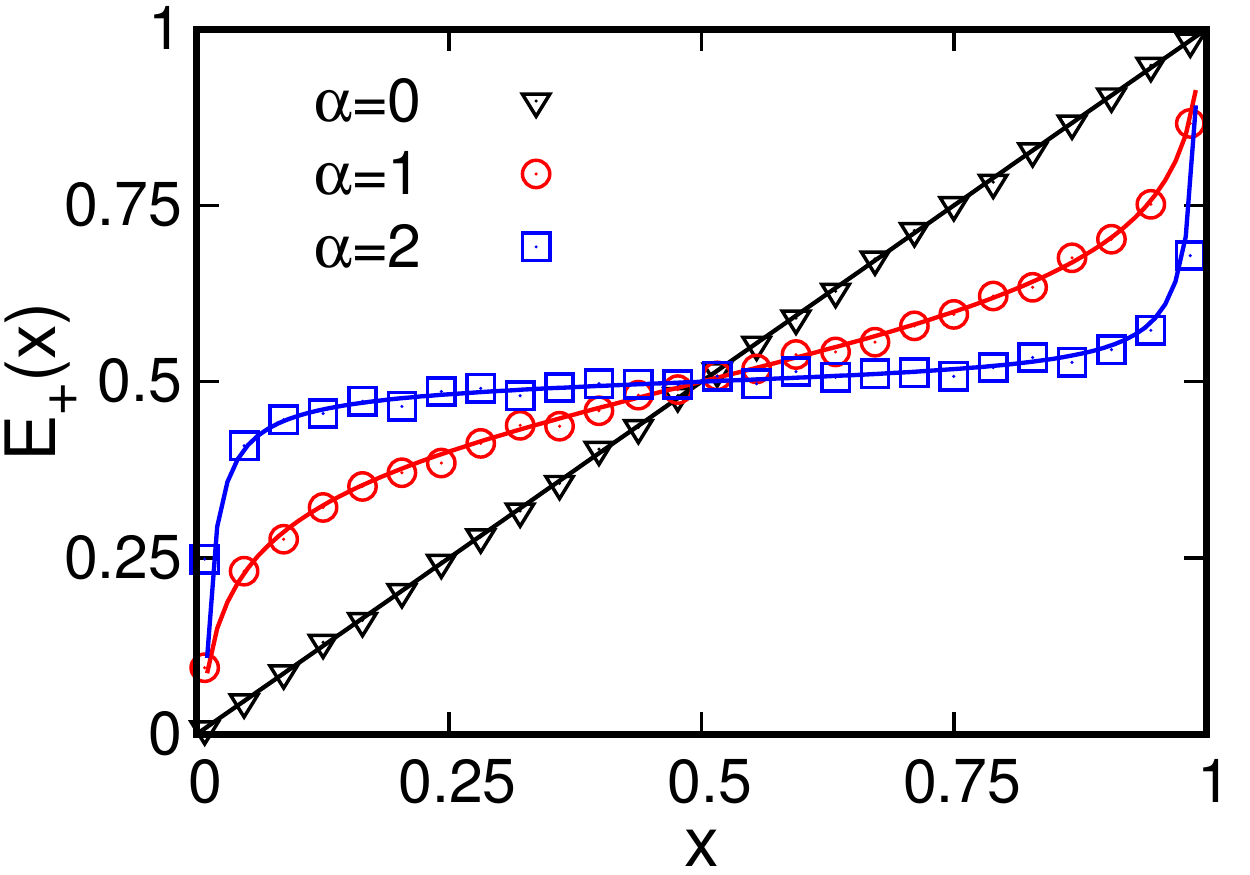}}
\caption{\small Exit probability versus the initial fraction of $+$ voters
  for the complete graph of 128 sites.  The solid curves represent
  Eq.~\eqref{Ep} and symbols give simulation results from $10^4$
  realizations.  To achieve the case $\alpha=1$, we choose $\ell=1$ and
  $p=1/2$, while $\alpha=2$ arises by choosing $\ell=1$ and $p=2/3$.}
\label{exit}  
\end{figure}

(ii) When the two news sources are equally connected to the population, the
polarization time $T_{\rm pol} $ scales as
\begin{align}
 \label{Tp2}
  T_{\rm pol}\sim N\,\frac{1-p}{p\ell}\,.
\end{align}
Hence, political polarization occurs quickly when voters are better connected
to competing news sources.  (iii) The exit probability has an anti-sigmoidal
shape (Fig.~\ref{exit}) because the competing news sources drive the
population to a politically polarized state. (iv) For the complete and the
two-clique graph, the opinion distribution undergoes a transition from a
homogeneous to a polarized state as the influence of news sources on voters
become stronger.

We now outline the calculations that underlie our results.  Suppose that we
know $r_{\pm}(x)$, the rates for $x$, the fraction of voters with $+$
opinion, to change by $\pm \frac{1}{N}\equiv \pm\delta x$.  Let
$P(x,t)\delta x$ be the probability that the fraction of $+$ voters lies
between $x$ and $x+\delta x$.  The Fokker-Planck equation for $P$ is
\begin{align}
  \frac{\partial P}{\partial t} =\mathcal{L}P\,,\quad \mathcal{L}= -\frac{\partial }{\partial x}V(x) + \frac{\partial^2 }{\partial x^2}D(x)\,,
 \label{FP}
\end{align}
with drift velocity $V(x)=[r_+(x)-r_-(x)] \delta x$ and diffusion coefficient
$D(x)=[r_+(x)+r_-(x)] \delta x^2/2$.  We can view the instantaneous opinion
$x$ as undergoing biased diffusion in the interval $[0,1]$ in the presence of
the effective potential
\begin{align}
  \label{potential2}
\phi(x)=-\int^{x} \frac{V(x')}{D(x')}\,dx'\,.
\end{align}

A basic opinion characteristic is the exit probability $E_+(x)$.  This
quantity satisfies the backward equation
$\mathcal{L}^{\dagger}E_+(x)=0$~\cite{G85,K97,R01}, where the adjoint
operator is
\begin{align}
 \mathcal{L}^{\dagger} \equiv
 V(x) \frac{\partial}{\partial x}+ D(x) \frac{\partial^2 }{\partial x^2}\,,
\end{align}
subject to the boundary conditions $E_+(0)=0$, $E_+(1)=1$.  The formal
solution for $E_+(x)$ is
\begin{align}
 \label{eqE}
 E_+(x)=
  \frac{{ \int^{x}_{0} \exp[\phi(x')]dx}}
  {{ \int_0^1 \exp[\phi(x')]dx'}}\,.
\end{align} 
By normalization, the fraction of trajectories that reach $x=0$ without
reaching $x=1$ is $E_-(x)=1-E_+(x)$.

Similarly, the consensus and polarization times satisfy the backward equation
$\mathcal{L}^{\dagger}T(x)=-1$~\cite{G85,K97,R01}.  The boundary conditions
for $T_{\rm con}$ are $T(0)=T(1)=0$, while the boundary conditions for
$T_{\rm pol}$ are $\frac{\partial T}{\partial x}\big|_{x=0}=0$ and
$T(\frac{1}{2})=0$.  The formal solutions are~\cite{BR20}
\begin{align}
  \begin{split}
    \label{TcTp}
     T_{\rm con}(x)&=E_+(x)I(x,1)- E_-(x)I(0,x)\, ,  \\
   T_{\rm pol}(x)&=I(x,1/2)\, ,
  \end{split}
\end{align}
where
$I(a,b)=\int^{b}_{a}dx'\int^{x'}_{0}dx''\exp[\phi(x')-\phi(x'')]/D(x'')$.

We now apply an annealed-link approximation to this formalism to determine
$E_+(x)$, $T_{\rm con}$, and $T_{\rm pol}$ for voters on the complete and the
two-clique graphs (Figs.~\ref{graphs}(a) and (b)).  In this
approximation, we replace the true transition rates for each voter on a given
fixed-link network realization by the average transition rate, in which a
link is present with probability proportional to its density.

\emph{Complete Graph:} By straightforward enumeration of all relevant events,
the transition rates $r_{\pm}(x)$ for voters on the complete graph are:
\begin{align}
 \label{eq1}
  \begin{split}
    r_+(x)&= \tfrac{1}{2} NAx(1-x) +B_+ (1-x) \,,\\
    r_-(x)&=\tfrac{1}{2} NAx(1-x)+B_-x\,.
\end{split}
\end{align}
The first term in $r_{\pm}$ accounts for a voter that adopts the opinion of a
neighboring voter and the second term accounts for adopting the opinion of
the news source.  The coefficients $A$ and $B\pm$ are
\begin{subequations}
  \label{AB}
\begin{align}
  A=&\frac{(1\!-\!\ell_+)(1\!-\!\ell_-)}{1\!-\!(1/N)}
      +\frac{(1-p)(\ell_++\ell_--2\ell_+\ell_-)}
      {(1-p)+(2p-1)/N}\nonumber \\
   &~~~~~~~~~+\frac{(1-p)\ell_+\ell_-}{(1-p)+(3p-1)/N}\,,
\end{align}
\begin{align}
  B_{\pm}=&\frac{p\ell_{\pm}}{2}\!
            \left[\frac{1\!-\!\ell_{\mp}}{(1\!-\!p)\!+\!(2p\!-\!1)/N}
            \!+\!\frac{\ell_{\mp}}{(1\!-\!p)\!+\!(3p\!-\!1)/N}\right].
\end{align}
\end{subequations}
Using \eqref{eq1} and \eqref{AB} in the definitions of $V(x)$ and $D(x)$,
their ratio is
\begin{align}
\label{VD0}
  \frac{ V(x)}{D(x)} &= \frac{2\left[B_+(1-x)-B_-x\right]}{Ax(1-x) + (1/N)\left[B_+(1-x)+B_-x\right]} ~.
\end{align}

Importantly, $V/D$ is of order 1, except when $x$ is of order $1/N$ away from
the boundaries at 0 and 1.  Within these boundary layers, the second term in
the denominator of $V/D$ ensures that $V/D$ remains finite even when $x=0,1$.
Considerable simplification arises by excluding these thin boundary layers
and consequently dropping this second term.  This approximation has a
vanishingly small effect on the consensus time for large $N$.  We find the
positions of the resulting slightly smaller interval $[a_-,1-a_+]$ by
equating the two terms of the denominator of $V/D$.  This gives
$a_\pm= {B_\mp}/{AN}$.  In this truncated interval, we have
\begin{align}
  V(x)=\frac{\left[B_+(1\!-\!x)-B_-x\right]}{N}  \quad  D(x)\approx \frac{Ax(1\!-\!x)}{2N} \,.
  \label{VD2}
\end{align}

Using this approximation for $V(x)$ and $D(x)$, the effective potential in
\eqref{potential2} becomes
\begin{align}
  \label{phi-eff}
  \phi(x)=-\ln [x^{\alpha_+}(1-x)^{\alpha_-}]\,,
\end{align}
where $\alpha_\pm=2B_\pm/A$.  We can also explicitly evaluate the integrals
in Eqs.~\eqref{eqE}--\eqref{TcTp} for specific values of $\alpha_{\pm}$.  For
simplicity, we specialize to the symmetric case of equally connected news
sources, so that $\alpha_+=\alpha_-=\alpha$, and correspondingly
$a_+=a_-\equiv a =\alpha/(2N)$.  Performing the integral in
Eq.~\eqref{potential2} with the potential in \eqref{VD2}, the exit
probabilities for $\alpha=1$ and $\alpha=2$ are (Fig.~\ref{exit})
\begin{align}
\label{Ep}
 E_+(x)=\frac{1}{2}\Big[1- \frac{H_\alpha(x)}{H_\alpha(a)}\Big]\, 
\end{align}
where
\begin{align*}
  H_{1}(x)&=\ln \left(x^{-1}-1\right)\\
  H_2(x)&=x^{-1}-(1-x)^{-1}+\ln \left(x^{-1}-1\right)^2\,.
\end{align*}
The anti-sigmoidal shape of $E_+(x)$ arises because the effective potential
\eqref{phi-eff} tends to drive the population to the politically polarized
state of $x=\frac{1}{2}$.

To obtain the consensus time, we substitute Eq.~\eqref{VD2} into the first
of Eqs.~\eqref{TcTp} to give
\begin{align}
\label{Tc}
 T_{\rm con}(x)=N\left[G_{\alpha}(a)-G_{\alpha}(x)\right] 
\end{align}
where, for simple rational values of $\alpha$, $G_\alpha$ is
\begin{align*}
  G_{\frac{1}{2}}(x)&=-4\arcsin \sqrt{x} \arcsin \sqrt{1-x}\,,\\
  G_{1}(x)&=-\ln \left[x(1-x)\right]\,,\\
G_{\frac{3}{2}}(x)&=(x\!-\!\tfrac{1}{2})\left[\arcsin \sqrt{x}\!-\!\arcsin
            \sqrt{1\!-\!x}\right]/\sqrt{x(1\!-\!x)}\,,\\
 G_{2}(x)&=\tfrac{1}{6}\left[x^{-1}(1-x)^{-1}-2\ln \left[x(1-x)\right]\right]\,.
\end{align*}
These give $T_{\rm con} \sim N $ for $\alpha=\frac{1}{2}$, $T_{\rm con} \sim N \ln N $ for
$\alpha=1$, $T_{\rm con} \sim N^{3/2} $ for $\alpha=\frac{3}{2}$ and $T_{\rm con} \sim N^2 $
for $\alpha=2$, as in Eq.~\eqref{Tc-N}.

\begin{figure}[ht]
\centerline{\includegraphics[width=0.35\textwidth]{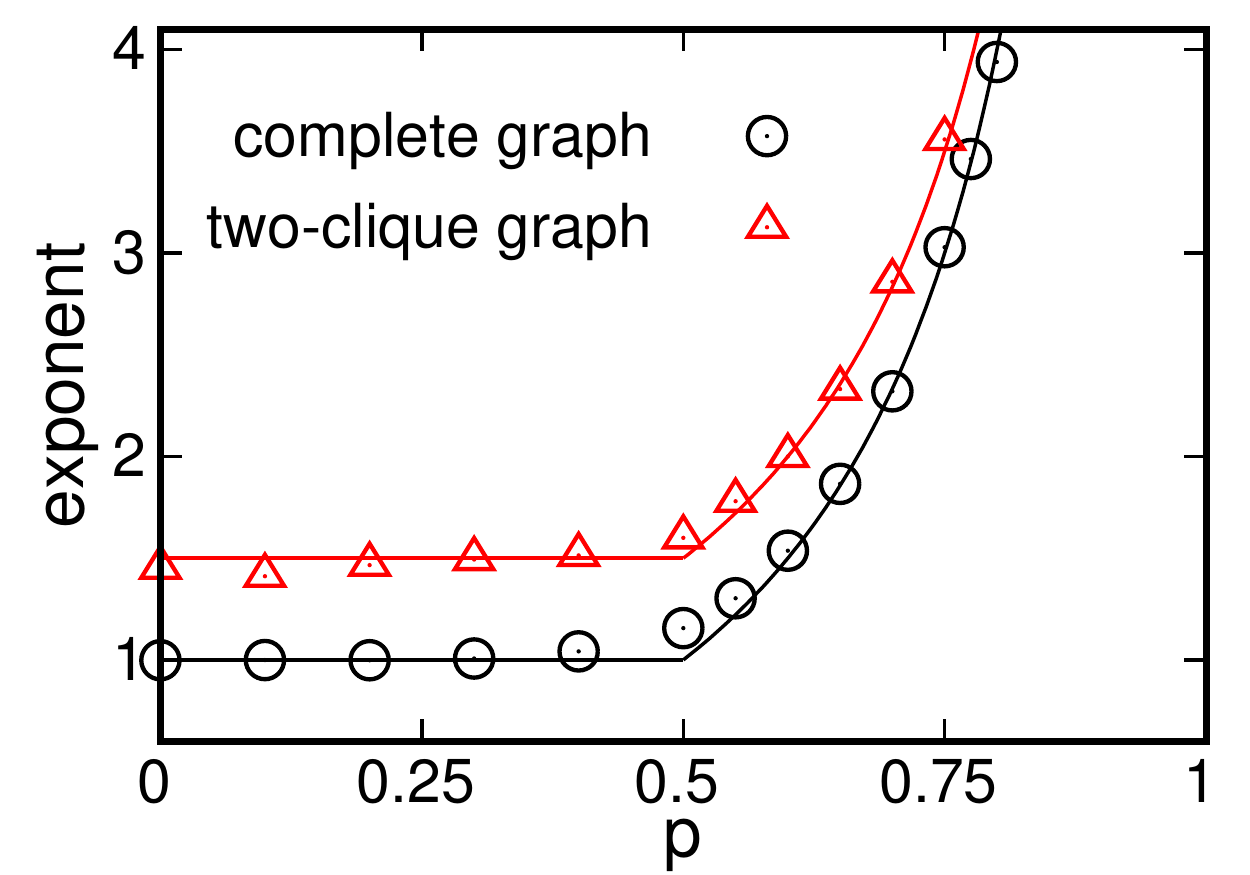}}
\caption{\small Consensus time exponents versus $p$ for link
  density to news sources $\ell=1$ and, for the two-clique graph, interclique
  link density exponent $\beta=\frac{1}{2}$.  Symbols are simulation results,
  while the curves are annealed-link approximations.}
\label{fig-exponent}  
\end{figure}

We can understand the $N$ dependence of $T_{\rm con}$ for arbitrary $\alpha$
in terms of the effective potential \eqref{phi-eff}.  According Kramers'
theory~\cite{K40}, the time to reach the boundaries at $a$ and at $1-a$ are
proportional to $\exp [\phi(a)]$ and to $\exp[\phi(1-a)]$, respectively.
Because the potential scales logarithmically in $N$ as $x\to a$ or
$x\to 1\!-\!a$, there is an algebraic, rather than an exponential, dependence
of $T_{\rm con}$ on $N$.  This behavior contrasts with voter models with
non-conserved dynamics~\cite{Lambiotte,Lambiotte3}, where the effective
potential leads to a consensus time that grows exponentially in $N$.  For
$\alpha<1$, the effect of the logarithmic potential is subdominant with
respect to fluctuations~\cite{Hirschberg} and the latter drive the system to
consensus, leading to $T_{\rm con} \sim N$.  These predictions agree with the
simulation results in Fig.~\eqref{fig-exponent}.  When $\ell_+\ne \ell_-$, the
lowest barrier height in the potential determines the exponent; therefore
$\alpha=\min(\alpha_+,\alpha_-)$ as in Eq.~\eqref{Tc-N}.  Finally, we numerically
verified that there is negligible difference in the consensus time when
connections between the two news sources and the population are random or
disjoint, with the same density of links.

To determine $T_{\rm pol}$ in a simple way, consider the extreme case where each news
source has a single link to the complete graph.  This weak connectivity leads
to the longest possible polarization time.  Suppose that the system starts in
the $-$ consensus state.  At some point, the ``informed'' voter, the one that
is linked to the $+$ news source, changes its opinion from $-$ to $+$.  When
this happens, this informed voter now disagrees with all its neighbors.  From
this excited state, subsequent opinion changes primarily occur among voters
within the complete graph.  Since there is only a single link to the news
sources, they play a negligible role in subsequent opinion changes.

\begin{figure}[ht]
\centerline{\includegraphics[width=0.275\textwidth]{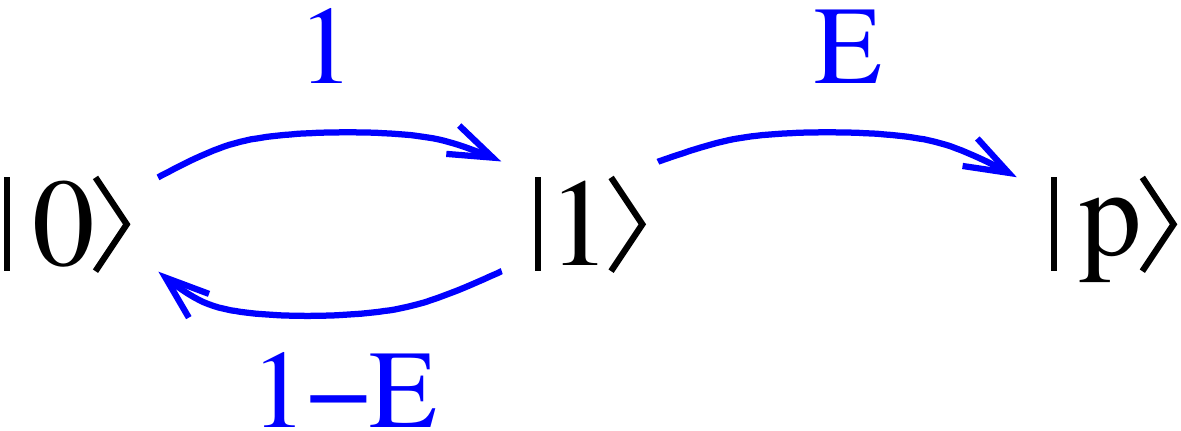}}
\caption{\small State space of the reduced system.}
\label{state}  
\end{figure}

The state space of this reduced system is schematically represented in
Fig.~\ref{state}.  Here $|0\rangle$ denotes the consensus state, $|1\rangle$
denotes the excited state where the informed voter has changed opinion, and
$|p\rangle$ denotes the polarized state in which the fraction of $+$ and $-$
voters are equal, and $E$ is the exit probability to reach $|p\rangle$, which
equals $\frac{2}{N}$~\cite{BR20}.  We can now write the following backward
equations for the polarization time
\begin{align}
  \label{T01}
  T_{\rm pol}=dt_0+T_{\rm pol}'\quad~ T_{\rm pol}' = (1\!-\!E)(dt_1+T_{\rm pol})+E\tau\,.
\end{align}
Here $T_{\rm pol}$ and $T_{\rm pol}'$ are the times to reach the polarized state starting
from the states $|0\rangle$ and $|1\rangle$, respectively,
$dt_0=1/[r_+(0)\!+\!r_-(0)]$ is the time to leave the state $|0\rangle$,
$dt_1\approx 1$ is the time to leave the state $|1\rangle$, and
$\tau=2N(1-\ln 2)$ is the conditional time to reach the state $|p\rangle$
from $|1\rangle$ by voter model dynamics~\cite{BR20}.  Solving these
equations gives Eq.~\eqref{Tp2}.  We emphasize that when the news sources are
well connected to the population, the polarization time $T_{\rm pol}$ is less than
the consensus time because for $T_{\rm pol}$ the state of the system is driven
towards the minimum of the effective potential, while for $T_{\rm con}$ the system
has to surmount a potential barrier.

Finally, we obtain the steady-state opinion distribution,
$P_{\rm ss}(x)\equiv P(x,t\rightarrow \infty)$, by setting
$\frac{\partial P}{\partial t}=0 $ in Eq.~\eqref{FP}.  We also need to apply
reflecting boundary conditions because for all $\alpha>0$, the endpoints are
not fixed points of the stochastic dynamics.  Imposing normalization, we find
\begin{align}
  P_{\rm ss}(x)=\frac{x^{\alpha_+-1}(1-x)^{\alpha_--1}}
 {{B}\left[1-a_+;\alpha_+,\alpha_-\right]-{B}\left[a_-;\alpha_+,\alpha_-\right]}\,,
\end{align}
where $B(x;y,z)$ is the incomplete beta function.  For $\ell_+=\ell_-=\ell$,
$P_{\rm ss}(x) \propto [x(1-x)]^{\alpha-1}$.  This distribution undergoes a
bimodal to unimodal transition as $\alpha$ passes through 1.

\emph{Two-clique graph:} We can adapt the above argument for the polarization
time on the complete graph to obtain both $T_{\rm con}$ and $T_{\rm pol}$ on sparsely
interconnected two-clique graphs, where $\beta\to 0$.  Because the fraction
of interclique links is negligible compared to intraclique links, the opinion
dynamics when opinions in a single clique are not unanimous reduces to that
of \emph{isolated} cliques that are additionally influenced by news sources.
Let $x_i$ be the fraction of $+$ voters on clique $C_i$
(Fig.~\ref{graphs}) and denote the state of the system by $(x_1,x_2)$.
It is convenient to take the initial condition as the maximally polarized
(MP) state $(1,0)$.  The population tends to remain close to the MP state
because: (a) news sources tend to drive opinions to this state, and (b) the
time $dt_0$, the inverse of the probability for a $\pm$ interaction between
voters, which scales as $N^{1-\beta}$, is large for $\beta\to 0$.

For an isolated clique connected to a single news source, we obtain the
probability to reach the state $x_1=0$ by setting $\ell_-=0$, $\ell_+=\ell$
for $V/D$ in Eq.~\eqref{VD0} and using the resulting form in
Eqs.~\eqref{potential2} and \eqref{eqE} to give
 \begin{align}
 E(x_1)= 
 \begin{cases}
  1- \frac{\left(\alpha+2Nx_1\right)^{1-\alpha}-\alpha^{1-\alpha}}{\left(\alpha+2N\right)^{1-\alpha}-\alpha^{1-\alpha}} &\qquad \alpha\neq 1\\[2mm]
   1-\frac{\ln \left(2Nx_1+1\right)}{\ln \left(2N+1\right)}& \qquad \alpha=1\,.
 \end{cases}
 \label{exit-amass}
\end{align}
Using the same argument as in Eq.~\eqref{T01}, where the MP state, the MP
state with one opinion change, and $-$ consensus correspond $|0\rangle$,
$|1\rangle$, and $|p\rangle$ respectively, we can compute $T_{\rm con}$ and
obtain Eq.~\eqref{two-clique-tc}.  A closely related argument gives
$T_{\rm pol}$ in Eq.~\eqref{Tp2}.

\begin{figure}[ht]
\centerline{  \includegraphics[width=0.35\textwidth]{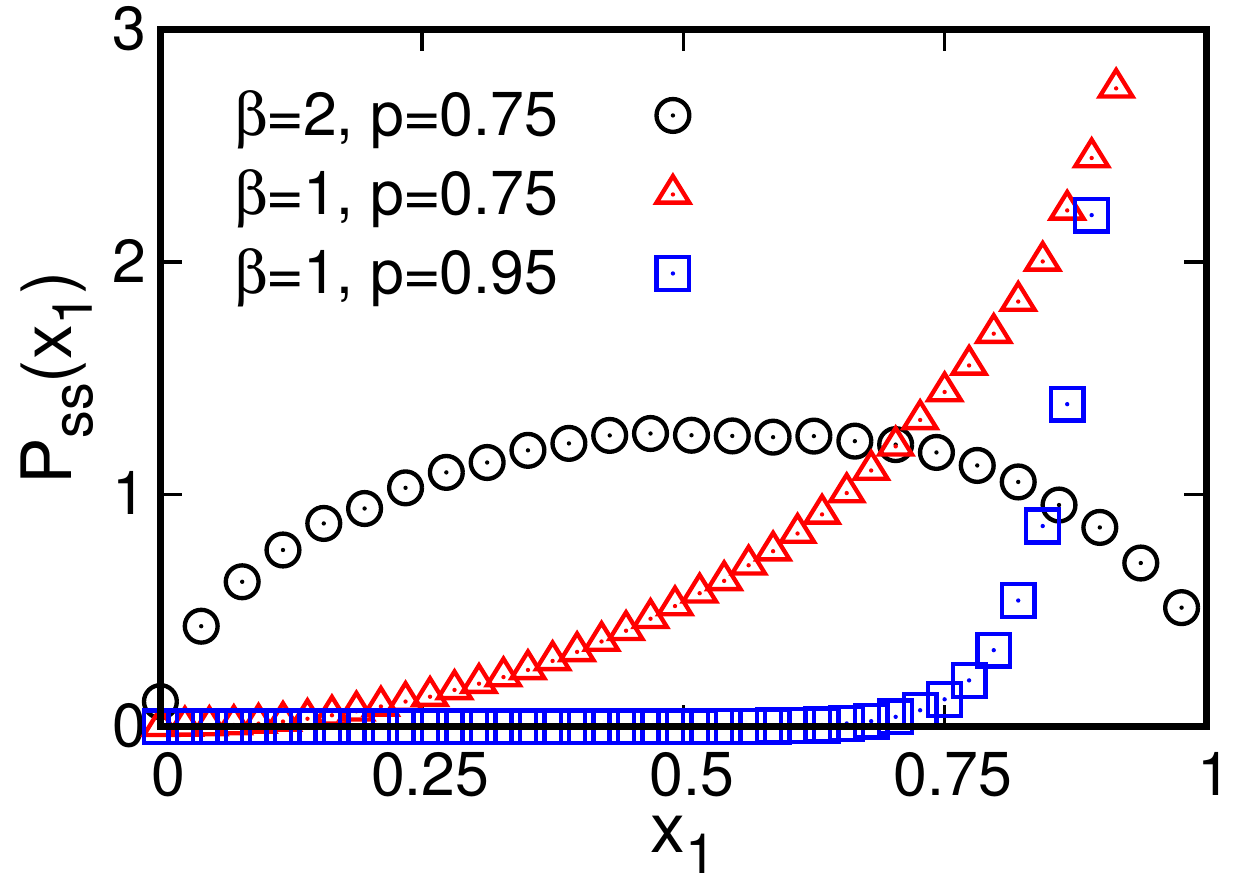}}
\caption{\small Distribution of fraction $x_1$ of $+$ opinion voters on
  clique $C_1$ of 128 voters on the two-clique graph, with $\ell=1$.}
\label{distribution-twoclique}  
\end{figure}
Finally, the steady-state distribution of $x_i$ normalized for each clique
(Fig.~\ref{distribution-twoclique}) shows that the opinions in the two
cliques indeed becomes more polarized as the number of interclique links is
reduced or the interactions with news sources become stronger.

To summarize, the presence of two well-connected antagonistic news sources
promotes political polarization in the voter model.  The news sources give
rise to an effective potential that leads to an anomalously long consensus
time and a short time to reach a politically polarized state.

\acknowledgements{We thank Mirta Galesic for inspiring discussions and
  gratefully acknowledge financial support from NSF grant DMR-1608211.}

\bibliographystyle{apsrev4-1}
\bibliography{reference}

\end{document}